\documentclass[conference]{IEEEtran}
\IEEEoverridecommandlockouts

\usepackage{cite}
\usepackage{amsmath,amssymb,amsfonts}
\usepackage{algorithmic}
\usepackage{graphicx}
\usepackage{textcomp}
\usepackage{xcolor}
\usepackage{hyperref}
\usepackage{multirow}
\usepackage{booktabs}
\def\BibTeX{{\rm B\kern-.05em{\sc i\kern-.025em b}\kern-.08em
    T\kern-.1667em\lower.7ex\hbox{E}\kern-.125emX}}
\begin{document}

\title{Predicting total time to compress a video corpus using online inference systems
\thanks{This work was funded by DTIF EI Grant No DT-2019-0068 for the Video Intelligent Search Platform (VISP) project.}
}

\author{\IEEEauthorblockN{Xin Shu}
\IEEEauthorblockA{\textit{Electronic and Electrical Engineering} \\
\textit{Sigmedia Group, Trinity College Dublin}\\
Dublin, Ireland \\
xins@tcd.ie}
\and
\IEEEauthorblockN{Vibhoothi Vibhoothi}
\IEEEauthorblockA{\textit{Electronic and Electrical Engineering} \\
\textit{Sigmedia Group, Trinity College Dublin}\\
Dublin, Ireland \\
vibhootv@tcd.ie}
\and
\IEEEauthorblockN{Anil Kokaram}
\IEEEauthorblockA{\textit{Electronic and Electrical Engineering} \\
\textit{Sigmedia Group, Trinity College Dublin}\\
Dublin, Ireland \\
anil.kokaram@tcd.ie}
}


\maketitle

\begin{abstract}
Predicting the computational cost of compressing/transcoding clips in a video corpus is important for resource management of cloud services and VOD (Video On Demand) providers. Currently, customers of cloud video services are unaware of the cost of transcoding their files until the task is completed. 
Previous work concentrated on predicting per-clip compression time, and thus estimating the cost of video compression.
In this work, we propose new Machine Learning (ML) systems which predict cost for the entire corpus instead. This is a more appropriate goal since users are not interested in per-clip cost but instead the cost for the whole corpus. In this work, we evaluate our systems with respect to two video codecs (x264, x265) and a novel high-quality video corpus. We find that the accuracy of aggregate time prediction for a video corpus is more than two times better than using per-clip predictions. Furthermore, we present an online inference framework in which we update the ML models as files are processed. A consideration of video compute overhead and appropriate choice of ML predictor for each fraction of corpus completed yields a prediction error of less than 5\%. This is approximately two times better than previous work which proposed generalised predictors.\end{abstract}


\section{Introduction}
\label{sec:introduction}
Video content delivery over the internet has surged since the COVID-19 pandemic.
Encoding and transcoding remain the heaviest computational tasks involved in video storage and streaming~\cite{YuriyReznik}. Predicting the compute load for compressing a video clip enables large-scale content delivery systems to allocate resources more efficiently, ensuring {\em task} completion within heterogeneous data centers. This also benefits cloud services, as prediction allows cloud customers to estimate transcoding/encoding costs before executing {\em tasks}~\cite{on_demand_cloud_video}. Currently, cloud video customers only learn the cost after the {\em task} is completed, making budgeting difficult without cost predictions. The challenge is that encoding time depends on visual content complexity~\cite{AnatoliyZabrovskiy}. Modern codecs (post H.264) are even more sensitive to content complexity since more elaborate processing is applied. Hence a clip of a few seconds containing high texture and motion might consume more compute cycles than one of a longer duration but with a stationary scene. 

Previous work~\cite{TewodorsDeneke,AnatoliyZabrovskiy,HadiAmirpour} uses encoding time as a proxy for compute load, and targets the prediction of encoding time per-clip.
However, users of cloud encoding services generally encode a large group of clips (a video corpus) at a time (e.g. television archives or production footage). In this case predicting the encoding time per clip is less important than predicting the encoding time for the entire corpus. We address this problem by developing a relatively simple set of ML predictors that adapt {\em as the corpus is processed}. This is therefore an instance of {\em online inference} in which our predictions update based on the content processed. 
An important advantage of this idea is that online inference eliminates the need to build particular models for different processing hardware, because the systems adapt to the hardware as processing proceeds. 
Overall we find that estimating the encoding time for a corpus improves prediction accuracy by more than 2 $\times$ compared with per-clip estimation. 
In addition, previous work reported accuracy in terms of log-time which does not give a true appreciation of the accuracy in compute load and budget predictions which are on linear scales. We report all our results in the linear domain. 
Furthermore, by choosing different predictors in relation to the percentage of the corpus that has been processed we can maximise performance over the whole range of operation.

Our contributions are as follows: 1) we build inference systems to predict corpus encoding time rather than per-clip encoding time and introduce a suitable metric {\em Sum Absolute Percentage Error (SAPE)} for evaluating performance; 2) we use Monte-Carlo sampling~\cite{monte_carlo} of our corpus to create different sampling sequences (100) to mitigate the impact of sequencing on our reported results; 3) we develop the idea of using different predictors at different corpus completion ratios to maximise performance.

\section{Background}
\label{sec:background}


 
Work on time-to-complete video encoding tasks has been conducted since 2015~\cite{TewodorsDeneke}. Deneke et al.~\cite{TewodorsDeneke} used a Multilayer Perceptron (MLP) Neural Network with a feature set that includes simple video properties/metadata. This set includes video resolution, duration, framerate, proportion of different frame types (I, P, B), and codec configurations. Zabrovskiy et al.~\cite{AnatoliyZabrovskiy} (H.264, H.265) in 2020 further introduced statistical spatial and temporal information along with video metadata as the feature set. With a similar neural network, they achieved average $R^2$ scores of 0.958~\cite{TewodorsDeneke} and 0.994~\cite{AnatoliyZabrovskiy} for predicting pre-clip transcoding time. Comparing these experiments, the results indicate the per-clip statistical features (Spatial and Temporal Information) as initially defined by ITU-T P.910~\cite{ITU_T_P910} are key features for achieving better accuracy.

More complex features, better correlated with compute time, were deployed by Menon et al.~\cite{pcs_caps} (H.265) and Amirpour (H.264, H.265) et al.~\cite{HadiAmirpour} derived from the energy of {\em Discrete Cosine Transform} coefficients used in compression. Again these were used to measure spatial and temporal energy defined as $E$ and $h$. In addition the average brightness (luma $L$) was also employed as a feature. Menon et al.~\cite{pcs_caps} deployed {\em XGBoost} regression algorithm~\cite{XGBoost} to achieve an $R^2$ score of 0.97. While Amirpour et al.~\cite{HadiAmirpour} further added a pre-trained {\em Convolutional Neural Network} (CNN) pre-processor to yield a latent vector for the 30th frame to the feature set. This vector was expected to contain rich information to enhance the prediction process. They achieved an $R^2$ score of 0.848 when predicting log-time using XGBoost. Note that in both of these cases, the test set was uniform in resolution and hence there was no video format information in the feature vector. Each study employed different encoders in testing and different compute instances which affects our ability to compare results.

All of the ML systems used in previous work were deployed as {\em generalised} predictors. The intention was that a single XGBoost or MLP model could be deployed to predict per-clip encoding time. Both Menon et al.~\cite{pcs_caps} and Amirpour et al.~\cite{HadiAmirpour} employed K-Means clustering on their training data so that an equal percentage of each cluster contributed to the final training/test set. This clustering step for organising the training data therefore allows the models to better generalise. But it also means that a new model has to be trained for every encoder/compute instance combination. We deploy this idea instead for online training of our models as processing progresses. Hence the entire corpus is clustered as previously before compression begins but the selection of clips for processing is informed by the cluster identity of each clip.

In general, all previous work has employed some form of ML system that takes a feature set as input to predict the encoding time of a clip. However, the corpus completion time as a more crucial target for processing a large group of clips has been overlooked. In this work, we employ a video corpus with clips of heterogeneous properties i.e. different framerates and duration, and each clip is from a different source. Hence we combine aspects of feature vectors from previous work and employ video metadata (video resolution, framerate, number of frames) as part of the feature vector. We adopt the {\em Video Complexity Analyser} (VCA) toolkit~\cite{VCA} to extract the DCT-energy-based features, $E$, $h$, and $L$. 
\section{Data and Experiment Setup}
\label{sec:dataset}
\begin{figure}[htb]
\begin{minipage}[b]{1.0\linewidth}
\centering
\centerline{\includegraphics[width=6.5cm]{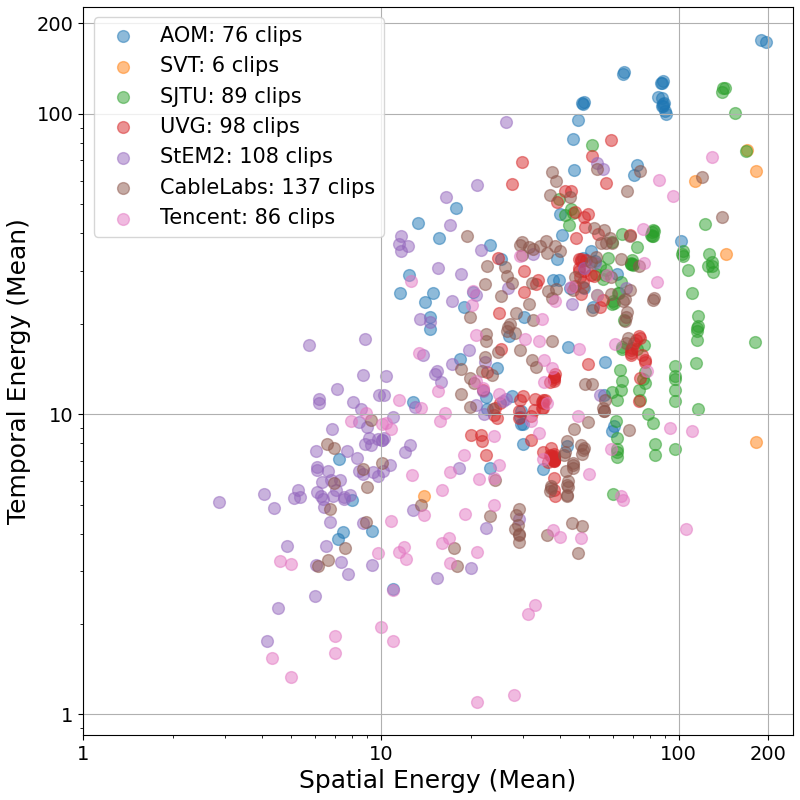}}
\caption{Scatter plot of the distribution of complexity features of the 600 segments. The x-axis indicates the Spatial Energy ($E$) and the y-axis indicates the Temporal Energy ($h$). The axes are transformed into logarithmic scales. The source datasets are labelled with colours.}
\label{fig:sete_distribuion}
\end{minipage}
\end{figure}

Our dataset contains 600 4K clips in raw format, which combines sequences employed in previous work \cite{AnatoliyZabrovskiy, VCD}, as well as new material from several other open-source datasets. These include AOM Common Test Conditions (AOM-CTC)~\cite{AOM}, SVT Open Content~\cite{svt_data}, SJTU 4K Sequences~\cite{SJTUdataset}, Tencent Video Dataset (TVD)~\cite{tvd_data}, Ultra Video Group dataset (UVG)~\cite{uvg_data}, CableLabs 4K Sequences~\cite{cablelabs}, and the StEM2 dataset from~\cite{smpte_data} the American Society of Cinematographers. The sequences included from~\cite{smpte_data} in particular have not been used before for this kind of study and we include them here to have a wide selection of high-quality material. All source sequences were 4K and normalised to YUV420p / 8-bit. The source material is segmented into short clips (2 or 4 secs) yielding our final set of 600 video clips. The distribution of complexity features ($E$, $h$) of the clips is visualised in Figure~\ref{fig:sete_distribuion}. Note that the axes are transformed into logarithmic scales for better visualisation. 
Unlike previous work, the source content of our 4-second segments differs from the 2-second segments. This removes duplication of content in our dataset and yields better coverage of spatiotemporal statistics. 

Encoding was conducted on a 2GHz AMD processor (EPYC-Rome 64C128T). We use Linux {\tt time} utility to measure the time (secs) required to complete each encoding task.
We consider two encoders: x264\footnote{\href{https://videolan.org/developers/x264.html}{\underline{x264 version 0.164}}, last access in April, 2024} and x265\footnote{\href{https://x265.readthedocs.io/en/stable/releasenotes.html}{\underline{x265 version 3.2.1}}, last access in April, 2024}. In each case we examine 12 combinations of encoding parameters which affect the time to compute: 3 {\em Presets} ({\em veryslow}, {\em medium}, and {\em ultrafast}) and 4 values of {\em Constant Quantization Parameters} (CQP) settings ($22$, $27$, $32$, and $37$), as recommended by the {\em Joint Video Experts Team} (JVET)~\cite{JVET_J1010}. To avoid CPU idling during the process and to standardise measurement, each clip was encoded with a single-threading configuration. Given these settings, we collect 14,400 samples of encoding time ($600 \times 2 \times 12$) for each encoding run over our 600 clip corpus.

\textbf{Data Transformation:} Previous work from Amirpour et al.~\cite{HadiAmirpour}, Menon et al.~\cite{pcs_caps} and Vibhoothi et al.~\cite{filling_the_gaps} shows that logarithmic transformation of encoding time reduces the internal covariate shift~\cite{internal_covariate_shift} phenomenon.
The log-transformed data are, therefore, more amenable to modelling and all our proposed systems are applied in this space. However, we report performance in linear space. This is an important aspect of our work. An error of 5\% in log space implies a much larger error in linear space, which is the space used for budgeting. 


\section{Proposed Systems}
\label{sec:algorithms}
Our systems are designed to estimate the per-clip encoding time, $t_i$, of the $i^{th}$ clip, and hence the aggregated encoding time $T$ for a corpus of $N$ clips. 
We define an {\em encoding task} as the compute task involved in encoding a single clip. We assume that a fraction $c$ of the corpus has been processed (i.e. completed). Hence $\hat{T}_{1-c}$ is an estimate of the encoding time for the fraction $(1-c)$ that has yet to be processed. In the sections that follow we present our ML and statistical predictors. The statistical predictors are based on relatively simple measurements derived from information available with a typical {\em progress bar} in some sense. These are therefore baseline systems that help us to assess the improvements using an ML based predictor.

\subsection{Statistical Predictors}
Our first Baseline Predictor (BP) of $t_i$, $\hat{t}^{BP}_i$ simply assumes the average encoding time for the fraction $c$ of clips is a good predictor of the time to encode each of the remaining fraction $1-c$. From this we can develop a predictor for the compute time for the rest of the corpus $T_{1-c}$, $\hat{T}^{BP}_{1-c}$, by scaling our estimate $\hat{t}_i^{BP}$ appropriately as follows.
\begin{align}
\hat{t}^{BP}_i = \frac{1}{c N} \sum_{k=0}^{c N}t_k;\ \ \ \ 
\hat{T}^{BP}_{1-c} = \frac{1-c}{c} \sum_{k=0}^{c N}t_k
\label{bp_predictor}
\end{align}
where $t_k$ indicates the recorded encoding time of the $k^{th}$ clip of the completed fraction ({\em cN}) of the corpus. 

We also develop an extension of our Baseline Predictor (BP), that uses statistical clustering to introduce some level of content adaptation in estimating $t_i, T_{1-c}$. This Clustering Predictor (CP) first assigns each clip $i$ to one of $J = 10$ clusters using K-Means clustering. The features for clustering are the video properties (height, number of pixels, framerate, and number of frames) and complexity features ($E, h, L$) discussed previously. This clustering step enables a different compute time prediction per content cluster hence better representing the remaining corpus. The average encoding time of completed encoding tasks $\hat{t}^{CP}_{i, j}$ (for cluster $j$) is then used to predict $t_{i, j}, T_{1-c}$ as follows.
\begin{align}
\hat{t}^{CP}_{i, j} = \frac{1}{cM_j} \sum_{k=0}^{cM_j} t_{k, j};\ \ \ \ 
\hat{T}^{CP}_{1-c} = (1-c)\sum_{j=0}^{J} M_j\ \hat{t}^{CP}_{i, j} \label{eq:cp_predictor}
\end{align} 
where each cluster $j$ has $M_{j}$ clips in total, and $t_{k, j}$ indicates the recorded encoding time of the $k^{th}$ clip of the completed fraction ({\em $cM_{j}$}) in cluster $j$. 
\subsection{XGBoost Predictors}
We develop two Machine Learning-based {\em online} inference systems based on XGBoost~\cite{XGBoost}. Given a feature vector $x_i$ and weights $w$ associated with a particular XGBoost model, we generate a prediction $f(x_i,w)$ for the compute time in log-scale for the $i^{th}$ clip. As discussed previously in this section, we actually examine compute time estimates, ($\hat{t}^{XGB}_i$), in linear-time and hence derive aggregated compute time measurements ($\hat{T}^{XGB}_{1-c}$) for the corpus as follows.
\begin{equation}
\begin{split}
\hat{t}^{XGB}_i &= \exp(f(x_i, w)) \\
\hat{T}^{XGB}_{1-c} &=\sum_{i=0}^{(1-c)N}\hat{t}^{XGB}_i=\sum_{i=0}^{(1-c)N}\exp(f(x_i, w))
\end{split}
\label{eq:xgb_predictor_aggregate}
\end{equation}
In our case, the input vector $x_i$ consists of the properties of the clips i) height, ii) number of pixels, iii) framerate, iv) number of frames, average per-frame values of the complexity features v) $E$, vi) $h$, vii) $L$, and encoder parameters viii) Preset and ix) CQP (constant quantisation parameter). We learn the model weights in log-time space hence our measured time values are transformed into log-space in training.

From this idea, we test two different variants of online XGBoost models in the context of BP and CP: i) online training using the fraction ($c$) of the corpus which is already processed (XP), ii) clustering the encoding tasks using K-Means followed by online training as previously (CXP).

In all of the {\em online} predictors presented (BP, CP, XP, CXP), the specific fraction of clips processed in the overall corpus will matter. There may exist a special fraction $c$ of the corpus which would yield a very good or very bad performance for any particular model. 
To mitigate this behaviour, we repeat the experiments with 100 Monte-Carlo~\cite{monte_carlo} realisations of data sampling from our corpus.
Our results therefore report the average performance of these realisations. 

\begin{table*}[htb]
\centering
\caption{Performance of systems presented. The values are taken from the average of 100 random sampling realisations of processing the corpus. The best performance of the metrics at each completion ratio is highlighted in bold.}
\label{tab:result_table}
\resizebox{\textwidth}{!}{%
\begin{tabular}{llrrrrrrrrrr}
\hline
\multirow{2}{*}{Inference Systems} & \multirow{2}{*}{Metrics} & \multicolumn{5}{c}{Completion ratio, $c$ (x264)} & \multicolumn{5}{c}{Completion ratio, $c$ (x265)} \\ \cmidrule(lr){3-7} \cmidrule(lr){8-12}
 & & @ 2\% & @ 6\% & @ 10\% & @ 20\% & \multicolumn{1}{l}{@ 40\%} & @ 2\% & @ 6\% & @ 10\% & @ 20\% & \multicolumn{1}{l}{@ 40\%} \\ 
 \cmidrule(lr){1-7} \cmidrule(lr){8-12}
\multirow{3}{*}{\begin{tabular}[c]{@{}l@{}}Baseline Predictor\\ (BP)\end{tabular}} & {\em MAPE} $\downarrow$ & 1074.68 & 1077.09 & 1095.81 & 1077.24 & 1100.79 & 1780.40 & 1799.22 & 1824.84 & 1830.30 & 1825.99 \\
 & $R^2$ $\uparrow$ & -0.02 & -0.01 & 0.00 & 0.00 & 0.00 & -0.01 & 0.00 & 0.00 & 0.00 & 0.00 \\
 & {\em SAPE} $\downarrow$ & 16.13 & 13.78 & 12.83 & 8.70 & 7.50 & 22.83 & 14.22 & 12.90 & 8.54 & 6.87 \\ \cmidrule(lr){1-7} \cmidrule(lr){8-12}
\multirow{3}{*}{\begin{tabular}[c]{@{}l@{}}Clustering Predictor\\ (CP)\end{tabular}} & {\em MAPE} $\downarrow$ & 1338.86 & 1126.16 & 1072.03 & 1096.19 & 1102.16 & 2239.63 & 1870.73 & 1790.03 & 1822.26 & 1831.94 \\
 & $R^2$ $\uparrow$ & -0.02 & 0.00 & 0.00 & 0.00 & 0.00 & -0.02 & 0.00 & 0.00 & 0.00 & 0.00 \\
 & {\em SAPE} $\downarrow$ & 21.84 & 6.18 & \textbf{4.89} & \textbf{3.26} & 3.06 & 23.02 & \textbf{6.98} & \textbf{5.11} & \textbf{3.83} & \textbf{3.51} \\ \cmidrule(lr){1-7} \cmidrule(lr){8-12}
\multirow{3}{*}{\begin{tabular}[c]{@{}l@{}}XGBoost Predictor\\ (XP)\end{tabular}} & {\em MAPE} $\downarrow$ & 32.87 & 26.02 & 23.65 & 21.30 & \textbf{19.42} & 38.91 & 30.73 & 27.91 & 24.95 & 22.12 \\
 & $R^2$ $\uparrow$ & 0.60 & 0.75 & 0.81 & 0.85 & \textbf{0.89} & 0.62 & 0.75 & 0.80 & 0.83 & 0.86 \\
 & {\em SAPE} $\downarrow$ & 17.92 & 9.38 & 6.73 & 4.26 & 2.98 & 19.42 & 10.62 & 7.75 & 5.33 & 3.97 \\ \cmidrule(lr){1-7} \cmidrule(lr){8-12}
\multirow{3}{*}{\begin{tabular}[c]{@{}l@{}}Clustering XGBoost\\ Predictor (CXP)\end{tabular}} & {\em MAPE} $\downarrow$ & \textbf{31.38} & \textbf{25.28} & \textbf{23.33} & \textbf{21.10} & 19.43 & \textbf{38.80} & \textbf{30.58} & \textbf{27.72} & \textbf{24.71} & \textbf{22.03} \\
 & $R^2$ $\uparrow$ & \textbf{0.76} & \textbf{0.80} & \textbf{0.82} & \textbf{0.86} & 0.89 & \textbf{0.73} & \textbf{0.79} & \textbf{0.81} & \textbf{0.84} & \textbf{0.87} \\
 & {\em SAPE} $\downarrow$ & \textbf{8.75} & \textbf{6.06} & 5.91 & 4.13 & \textbf{3.02} & \textbf{10.19} & 7.24 & 6.94 & 5.15 & 3.88 \\ \cmidrule(lr){1-7} \cmidrule(lr){8-12}
\multirow{3}{*}{\begin{tabular}[c]{@{}l@{}}Generalised XGBoost \\ Predictor (GXP)\end{tabular}} & {\em MAPE} $\downarrow$ & \multicolumn{5}{c}{35.54} & \multicolumn{5}{c}{42.98} \\
 & $R^2$ $\uparrow$ & \multicolumn{5}{c}{0.68} & \multicolumn{5}{c}{0.66} \\
 & {\em SAPE} $\downarrow$ & \multicolumn{5}{c}{13.37} & \multicolumn{5}{c}{19.17} \\ \hline
 \multirow{3}{*}{\begin{tabular}[c]{@{}l@{}}Generalised DNN \\ Predictor (GDP)\end{tabular}} & {\em MAPE} $\downarrow$ & \multicolumn{5}{c}{36.13} & \multicolumn{5}{c}{38.07} \\
 & $R^2$ $\uparrow$ & \multicolumn{5}{c}{0.65} & \multicolumn{5}{c}{0.68} \\
 & {\em SAPE} $\downarrow$ & \multicolumn{5}{c}{13.47} & \multicolumn{5}{c}{18.86} \\ \hline
\end{tabular}}
\end{table*}
\subsection{Generalised Predictors}
We also examine two generalised predictors that use Deep Neural-net (GDP) and XGBoost (GXP) as comparison for the {\em online} inference systems. They represent our effort to reproduce systems proposed in previous work: GDP as in ~\cite{AnatoliyZabrovskiy} and GXP as in ~\cite{HadiAmirpour}. 
GDP employs 22k neurons with 12 fully-connected layers and {\em LeakyReLU} activation function~\cite{LeakyReLU}. 
We do not use this DNN for online inference because the compute cost of training is higher than XGBoost (greater than 20$\times$) and would have to be repeated continuously as $c$ increases.
The test set for both GXP and GDP consists of the clips from {\em Tencent} and {\em CableLabs} dataset while the other sets (see Figure~\ref{fig:sete_distribuion}) constitute the training set. By examining the generalised predictors in the context of XP and CXP we can show the impact of clustering with training and the usefulness of online inference in practice. 




\section{Results and Discussion}
\label{sec:results}
We adopt 2 common metrics {\em Mean Absolute Percentage Error} ({\em MAPE}) and {\em Coefficients of Determination} ($R^2$) for assessing the per-clip time prediction as in previous work. We present the {\em Sum Absolute Percentage Error} ({\em SAPE}) for assessment of the aggregated encoding time prediction for processing the rest of the corpus as follows.
\begin{align}
SAPE &= \frac{\left |\sum_{i=0}^{(1-c)N} t_i - \sum_{i=0}^{(1-c)N} \hat{t}_i \right |}{\sum_{i=0}^{(1-c)N} t_i} \times 100\% \label{eq:eval_equations}
\end{align}
where $t_i$ is the actual encoding time of the $i^{th}$ clip, and $\hat{t}_i$ is the corresponding prediction. We assess the online inference systems with the average performance across 100 realisations at each 2\% increment of $c$ ranging from 2\% to 100\%. 

Detailed values of metrics at key intervals are listed in Table~\ref{tab:result_table}. As expected, all systems exhibit better accuracy in aggregated encoding time prediction for the corpus (SAPE) than per-clip time prediction (MAPE, $R^2$). Aside from the generalised predictors, all online systems perform better as $c$ increases. In general, the generalised predictors (GXP and GDP) underperform compared to the online inference systems (XP, CXP) in any metrics once the online inference starts, i.e., $c \geq 2\%$.

The compute cost is mainly due to calculations of complexity features ($E, h, L$) as well as updating the online inference engine when new transcoding tasks are completed. Inference cost is negligible. Our measurements show that the average overhead on 4K video segments is 0.17\% and 0.19\% respectively for 2-second and 4-second segments across the 12 encoder parameter combinations. With this in mind, we can rank the system complexities as follows from lowest to highest: GXP, GDP, BP, CP, XP, CXP. The systems other than GXP and GDP all require the transcoding to actually start, for example, BP is the classic progress bar predictor. Hence the online systems are more computationally complex than GXP and GDP, which can generate inferences before transcoding begins. Our results indicate that very high prediction accuracy can be generated by cascading predictors in the workflow. We therefore propose the following: GXP (x264) and GDP (x265) are the best predictors before transcoding begins; once transcoding begins, CXP is better than a progress bar (BP) up till 6\% of the corpus has been processed; thereafter CP is marginally better. In all cases, we can perform better than the simple progress bar.
\section{Conclusion}
\label{sec:conclustion}
We examined four {\em online} systems and two generalised systems aimed at estimating the aggregated compute time required for encoding a video corpus. We introduced the Sum Absolute Percentage Error {\em SAPE} to better assess such performance. Each of the systems have different useful domains of operations. Before the beginning of the corpus compression, GXP can provide a rough prediction at around 13.47\% (x264) / 18.86\% (x265) accuracy. Introducing CXP as an online inference system after 2\% of the corpus has been processed improves accuracy to 8.75\% (x264) / 10.19\% (x265). The system can then switch to CP after 6\% completion to further improve accuracy to 4.89\% (x264) / 5.11\% (x265) and at the same time reduce computational overhead. 

In this paper, we discover that a basic statistical model (BP, CP) can be as accurate as a generalised model in predicting the aggregated time after learning from a small fraction of the corpus. The accuracy can be further improved by introducing a low-cost {\em online} Machine-Learning technique (XP, CXP). This enables accurate compute cost/budget prediction for processing a large scale video corpus. In future work, we explore the use of simpler video content characteristics that incur even lower compute overhead. 

\newpage
\bibliographystyle{IEEEtran}
\bibliography{refs}

\end{document}